\documentclass{pnastwo}

\usepackage{bbm}

\newcommand{\mh}{\mathcal{H}}

\newcommand{\mm}{\mathcal{M}}
\newcommand{\rd}{\mathrm{d}}

\newcommand{\lp}{\left(}
\newcommand{\rp}{\right)}

\begin{document}
\title{Architecture and Co-Evolution of Allosteric Materials}
\author{Le Yan\thanks{lyan@kitp.ucsb.edu}\affil{1}{Kavli Institute for Theoretical Physics, University of California, Santa Barbara, CA 93106, USA}, Riccardo Ravasio\thanks{riccardo.ravasio@epfl.ch}\affil{2}{Institute of Physics, EPFL, CH-1015 Lausanne, Switzerland}, Carolina Brito\thanks{cacabrito_kk@yahoo.com.br}\affil{3}{Instituto de F\'isica, Universidade Federal do Rio Grande, do Sul CP 15051, 91501-970 Porto Alegre RS, Brazil}, Matthieu Wyart\thanks{matthieu.wyart@epfl.ch}
\affil{2}{Institute of Physics, EPFL, CH-1015 Lausanne, Switzerland}}

\contributor{Submitted to Proceedings of the National Academy of Sciences
of the United States of America}    

\significancetext{In allosteric proteins, binding a ligand  affects   activity at a distant site. The physical principle allowing for such an action at a distance are not well understood. Here  we introduce a numerical scheme to evolve allosteric materials in which the number of solutions, their spatial architectures and the correlations among them can be computed. We show that allostery in these materials uses recently discovered elastic edge modes near the active site to transmit information, and that correlations generated during evolution  alone can reveal key aspects of this architecture.}




\maketitle
\begin{article}

\begin{abstract}
We introduce a numerical scheme to evolve functional materials that can accomplish a specified mechanical task. In this scheme, the number of solutions, their spatial architectures and the correlations among them can be computed. As an example, we consider an ``allosteric'' task, which requires the material to respond specifically to a stimulus at a distant active site.  We find  that functioning materials evolve a less-constrained trumpet-shaped region connecting the stimulus and active sites,  and that the amplitude of the elastic response  varies non-monotonically along the trumpet. As previously shown for some proteins, we find that correlations appearing during evolution alone are sufficient to identify key aspects of this design.
Finally, we show that the success of this architecture stems from the emergence of soft edge modes recently found to appear near the surface of marginally connected materials.  Overall, our in silico evolution experiment offers  a new window to study the relationship between structure, function and  correlations emerging during evolution.

\end{abstract}


Proteins are long polymers that can fold in a reproducible way and achieve  a specific function. Often, the 
activity of the main functional site depends on the binding of an effector on a distant site \cite{Changeux05}. Such an allosteric  behavior   can occur over large distances, such as 20 residues or more \cite{Daily07}, and often  involves only a sparse subset of residues in the protein \cite{Daily07,mclaughlin12}. Allosteric regulation offers an appealing target for drug design \cite{Nussinov13}, 
and there is a considerable interest in predicting allosteric pathways \cite{Amor16,Halabi09}. One central difficulty is that the physical mechanisms allowing such an action-at-a-distance remain elusive. In some cases, allostery can be understood as the modulation of a hinge connecting two extended rigid parts of the protein \cite{Perutz70,Gerstein94}, but often the displacement field induced by the binding of the effector cannot be described in these terms \cite{Goodey08,Gandhi08,mclaughlin12}. Another route, statistical coupling analysis \cite{Lockless99}, considers correlations within sequences of proteins of the same family to infer  allosteric pathways \cite{mclaughlin12,Halabi09}. The generality of this elegant approach is however debated \cite{Tecsileanu15}.

\begin{figure}[htbp]
\centering
\includegraphics[width=.8\columnwidth]{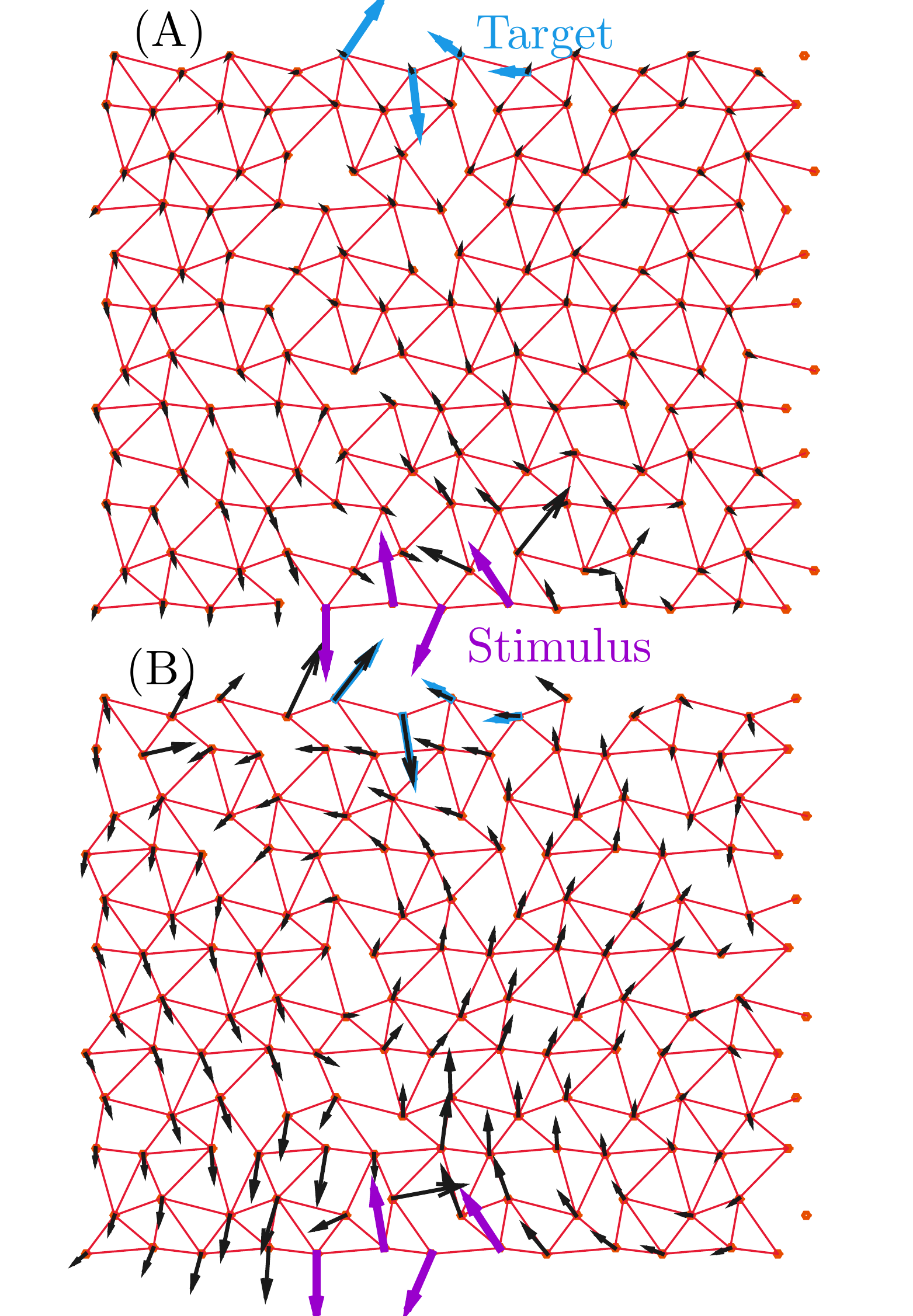}
\caption{
Illustration of the model. A network is defined by the location of strong springs (in red), connecting the adjacent nodes of a distorted triangular lattice. A network is fit if its response field (black arrows) to an imposed stimulus (purple arrows) reproduces a target displacement (cyan arrows) at a distant site, consisting here of four nodes. In (A), a random configuration ($T_e=\infty$) performs poorly.  In (B), the system has evolved by moving strong springs, and performs almost perfectly.  Here $T_e=0.01$, $L=12$ and  $z=5.0$. \label{model}}
\end{figure}

From a physical viewpoint, specific response at a distance is surprising. The structure of proteins is similar to randomly packed spheres  \cite{Liang01}. Generically, the response of such systems is non-specific and decays rapidly in space  (in a manner similar to a continuum medium) at distances larger than the particle size. This is true except close to a critical point where the number of constraints coming from strongly interacting particles is just sufficient to match the number of degrees of freedom of the particles \cite{Liu10}. There, the elastic response becomes heterogeneous on all scales   \cite{During13,Lerner14}. This point is illustrated in Fig.~\ref{model}.A showing the rapidly-decaying response of a random spring network to a stimulus. However, as shown in Fig.~\ref{model}.B (and independently found in \cite{Rocks16} using a different algorithm), springs can be moved so that the response extends further and specifically matches a target response on the other side of the system. This observation raises various questions, including: (i) Which network architectures allow for such allosteric response?  (ii) Why are such architectures working from a physical viewpoint? (iii) What is the number of solutions? (iv) What are the correlations among them, and are these correlations sufficient by themselves to identify key aspects of the architecture, as proposed for real proteins? 

In this work, we answer these questions by introducing a model of elastic networks that can evolve according to some fitness function $F$, which depends on the  response of the material to a well-defined stimulus. Our approach allows for a considerable freedom in the choice of the fitness function. As an illustration, we impose here that a displacement of four nodes on one side of the material (the ``stimulus") elicits a given displacement of identical amplitude but different direction on four target nodes on the other side of the system. A key advantage of our scheme is that our algorithm uniformly samples the fitness landscape (we use a Monte Carlo algorithm that turns out to equilibrate rapidly), which allows us to count the number of solutions and compute the entropy $S(F)$, as well as to guarantee that the solutions generated are the typical (most numerous) ones. The quality of the solutions can be monitored by an ``evolution temperature" $T_e$ that controls the fitness of the solutions probed. Our central findings are that: (i) there exists a transition temperature below which high-quality solutions appear and above which solutions are poor. (ii) High-quality solutions share a specific design. They present a trumpet-shaped region where the material is less constrained, which end by a marginally-constrained region in the vicinity of the target. (iii) The response amplitude varies non-monotonically between the stimulus and active sites. (iv) We rationalize this design based on a recent theory of edge modes in marginally connected disordered media  \cite{Yan16b}.  (v) We show that co-evolution --- the correlations in the structures of the family of solutions --- alone is sufficient to identify the trumpet structure. Finally, this detailed characterization of the solutions also points to some of their limitations in using them in thermal environments. We discuss how the fitness function can be changed to alleviate such problems. 

\section*{Description of the evolution model}

Scalar models, where the response of a node is described by a scalar instead of a vector, have been introduced to study co-evolution and allostery \cite{Hemery15,Tlusty16}. Although these models can capture the rigid motion of a part of the system, they cannot address the propagation of more complex mechanical information, such as that illustrated in Fig.~\ref{model}. Instead, we use elastic networks, which  have been used extensively to describe the vibrational dynamics of proteins \cite{Atilgan01,Rios05}.  

Specifically, we consider on-lattice models previously used to describe covalent glasses~\cite{Yan14,Yan15b}.  $N=L^2$ nodes are located on a triangular lattice (slightly distorted  to avoid  straight lines of nodes, see discussion in S.I.). $N_{\rm s}$ strong springs of stiffness $k_{\rm s}$ connect  some of the adjacent nodes, modeling strong interactions such as peptide bonds, disulfide bridges or other electrostatic interactions. We declare that  $\sigma_{\alpha}=1$ if a strong spring is present in the link $\alpha$ (as represented  in red in Fig.~\ref{model}), and $\sigma_{\alpha}=0$ otherwise. To ensure that the elastic response is always uniquely defined, each node also interacts with all its next nearest neighbors via weak springs of stiffness $k_{\rm w}=10^{-4}$,  mimicking   weaker interactions. Thus the network is entirely described by a connection vector $|\sigma\rangle$ whose dimension is the number of links. We define the average coordination number $z$ as $2 N_{\rm s}/N$.  Marginally connected networks corresponds to $z_c=4$ in two dimensions \cite{Maxwell64}, and we denote $\delta z=z-z_c$.

For a given configuration $|\sigma\rangle$, we consider the response  to a  ligand binding event, which we model as an imposed displacement field $|\delta{\bf R}^{\cal E}\rangle$ on a set of 4 adjacent nodes (the ``allosteric" site) located on one free boundary of the system, as illustrated in purple  in Fig.~\ref{model}. After relaxing the elastic energy, such a stimulus will generate a displacement field $|\delta{\bf R(\sigma)}^r\rangle$ in the entire system; a fast numerical calculation of this response is formulated in Methods. Here, we focus on studying networks for which the response generates a desired target displacement   $|\delta{\bf R}^{\cal T}\rangle$ of identical amplitude but different direction (illustrated in cyan in Fig.~\ref{model}) on an ``active'' site, which we also choose to consist of $n_{\cal T}=4$ nodes on the other side of the system. 
 Because physically only the strain (and not the absolute displacement) at the active site can affect catalytic properties, the target displacement is defined modulo a global translation and rotation $|{\cal {\bf U}}\rangle$. To rank networks in term of their allosteric ability, we define a fitness function $F$ and a cost $E$: 
\be
F(\sigma)\equiv-E(\sigma)\equiv -\min_{\cal |{\bf U}\rangle}\sqrt{\sum_{i\in{\cal T}}(\delta{\bf R(\sigma)}^r_i-\delta{\bf R}^{\cal T}_i-{\cal {\bf U}}_i)^2},
\label{cost}
\ee
where $i$ label the nodes, and ${\cal T}$ the set of nodes belonging to the active site. Thus $F=0$ corresponds to a perfect allosteric response. Below we restrict the networks further by imposing that all adjacent active nodes are connected, and choose a  target displacement that does not stretch these bonds. The minimization of Eq.(\ref{cost}) can be readily performed and the fitness can be written directly in terms of $|\delta{\bf R}^{\cal T}\rangle$ and $|\delta{\bf R(\sigma)}^r\rangle$, as discussed in Methods. 

\begin{figure}[htbp]
\centering
\includegraphics[width=1.\columnwidth]{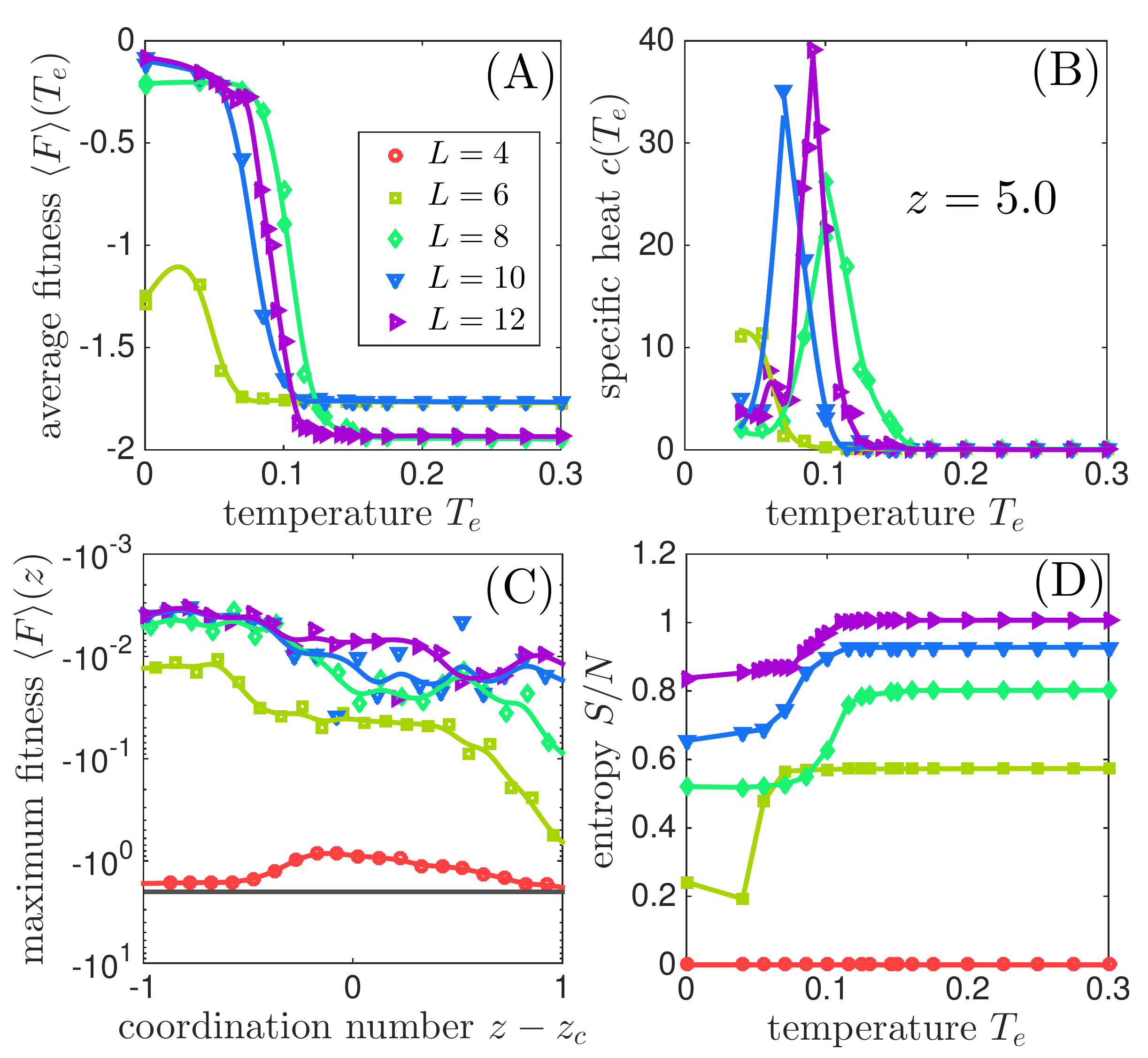}\\
\caption{  
A. Average fitness $\langle F\rangle$ versus evolution temperature $T_e$ for various system sizes where $z=5.0$. A steep change in fitness is seen near $T_c\approx0.09$ for relatively large systems. 
B.  Specific heat $c$ versus temperature $T_e$. The maximal specific heat increases with the system size $L$, suggesting the existence of a thermodynamic transition at $T_c$.   
C. Fitness averaged over local maxima $\langle F\rangle_{T_e=0}$ versus coordination number $z$ in log-linear scale. 
The black line shows the fitness if no mechanical response is present at the active sites. 
D. Entropy density $S/N$ versus temperature $T_e$. The entropy jump near $T_c$ indicates the number of degrees of freedom that must be tuned to achieve the desired response. 
\label{thermo}}
\end{figure}

Henceforth we consider two dimensional networks with periodic boundaries in the direction transverse to the direction between the allosteric and active sites (this corresponds to a cylindrical geometry). 
As is the case for many aspects of the microscopic elasticity of amorphous materials \cite{Liu10}, we  expect our results to hold independently of the spatial dimension.

\section*{Numerical solutions}
To evolve networks we  use a  Metropolis algorithm (see Methods) at some ``evolution temperature'' $T_e$, where strong springs can swap from an occupied to an unoccupied link, leaving the average coordination $z$ constant. More precisely, $T_e$ is the variable conjugate to the fitness, so that once our algorithm reaches equilibrium (which it does in the range of $T_e$ values presented here), the probability $P(\sigma)$ of finding a configuration $|\sigma\rangle$ reads  $P(\sigma)=\exp(F(\sigma)/T_e)/{\cal Z}$ where ${\cal Z}=\sum_\sigma \exp(F(\sigma)/T_e)$.  $T_e=\infty$ corresponds to random networks.

Thus as we lower the temperature, we probe fitter and fitter networks as illustrated in Fig.~\ref{model}. This point is systematically studied in Fig.~\ref{thermo}.A showing   $\langle F\rangle(T_e) $ for a given coordination, where the ensemble average is made on both Monte-Carlo steps and different realizations. 
For $T_e>0$, we find that this average does not depend on the time of the simulation if it is long enough, indicating that an equilibrium was reached. %
As $T_e$ decreases, we observe a transition from low to high fitness, which appears to become sharper and sharper as $N$ increases. This suggests a thermodynamic transition, as we evidence further by considering the specific heat $c(T_e) = \frac{\rd \langle E\rangle(T_e)}{\rd T_e}$ which displays a more and more pronounced peak as $L$ 
increases, as shown in Fig.~\ref{thermo}.B. This result is a signature of an underlying collective phenomenon, indicating that achieving an allosteric function is a collective process.

Below the transition, we find that the networks perform well. Their performance can be quantified by $\langle F\rangle|_{T_e=0}$,  an ensemble average of local maxima in the fitness landscape, as reported in Fig.~\ref{thermo}.C. These structures result from a pure gradient ascent in the fitness landscape. We find that in the range of coordination we probed, the cost decreases by at least $200$ folds with respect to random networks, i.e. the response converges very precisely toward the desired one. Thus, the systems does not get stuck in local maxima of poor quality in the fitness landscape. 

Finally, we can quantify the number of allosteric networks. It follows $e^{S(T_e)}$, where  $S(T_e)$ is the entropy. It satisfies
$\rd S= c(T_e)\frac{\rd T_e}{T_e}$, and is shown in Fig.~\ref{thermo}.D. For example, at $T_e=0.05$ where networks perform very well, we find that their number is very large -- exponential to the system size: $e^{S(T_e)}\approx10^{53}$ for $L=12$, but the probability $p_A$ to obtain such a network by chance is also exponentially low: $p_A=e^{S(T_e)-S(\infty)}\approx10^{-10}$ for $L=12$. 

\section*{Architecture  of allosteric networks}
Hypostatic networks with $\delta z<0$ are extremely floppy. It may be a interesting case to study intrinsically disordered  proteins \cite{Dunker08}, but for folded proteins considering  $\delta z>0$ is more realistic. Henceforth we focus on that case and choose $z=5$. In S.I. our results are presented for the floppy case $\delta z<0$. 

\begin{figure}[htbp]
\centering
\includegraphics[width=1.\columnwidth]{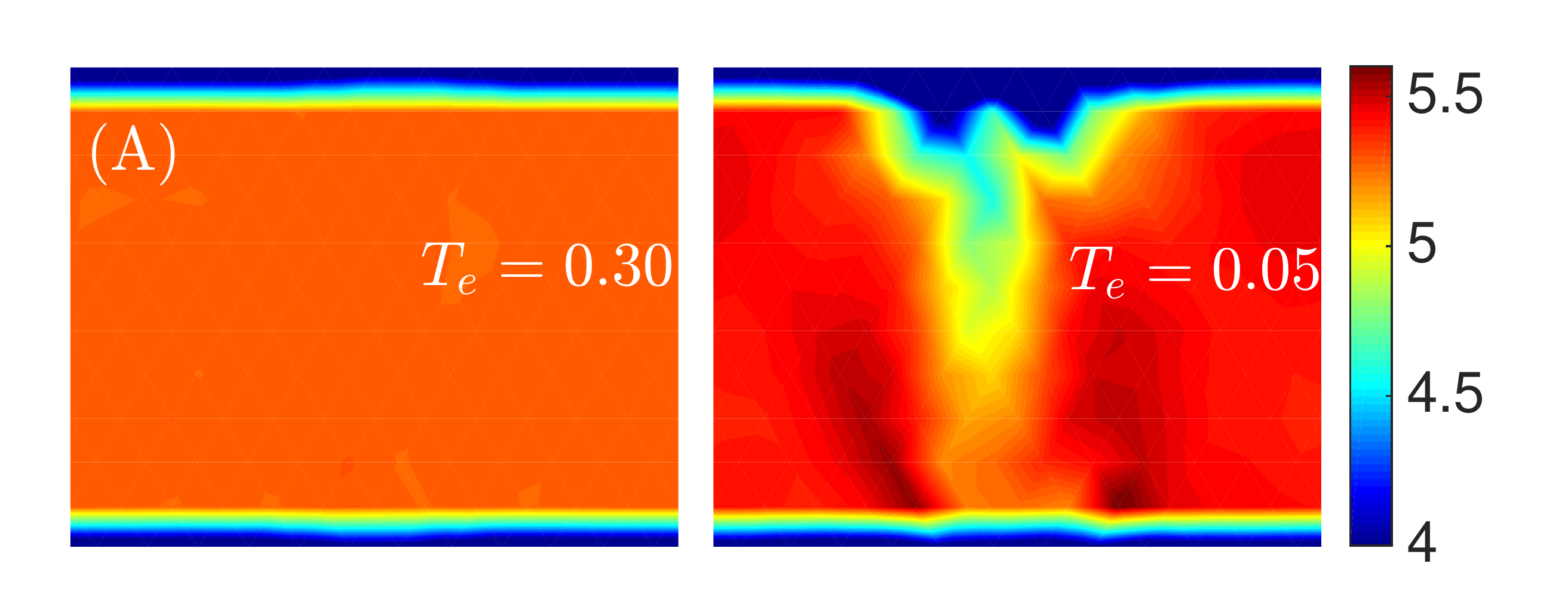}\\
\includegraphics[width=1.\columnwidth]{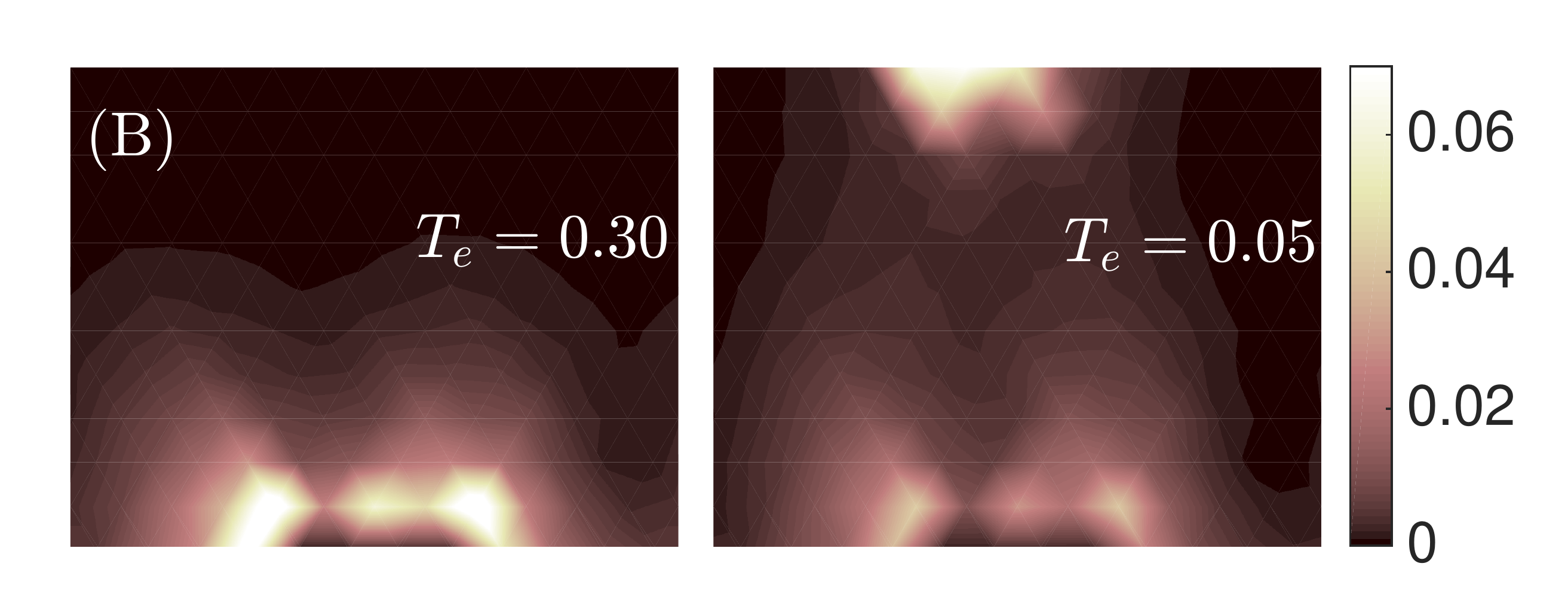}
\caption{ 
A. Map of the  mean coordination number and 
B. Spatial distribution of the average response magnitude for configurations equilibrated at 
 $T_e=0.30$ (left) and $T_e=0.05$ (right). 
 In the functioning networks (right), a trumpet connecting the allosteric and active sites appear in A, and the response to stimulus varies non-monotonically inside the trumpet in B.
\label{zp}}
\end{figure}

Which architecture allows for such a long-distance, specific  response?
A systematic design is revealed by averaging the occupancy of various solutions which our algorithm generates, as shown in Fig.~\ref{zp}.A. At high temperature,  the structures are essentially random and not functioning. At low temperature, a trumpet-shaped region appears that connects the allosteric and active sites. Specific features are that:
\begin{itemize}
\item Inside the trumpet, the mean occupancy is lower than the mean, but there are no floppy modes (i.e. modes that do not deform the strong springs). 
\item The mean occupancy or coordination decreases monotonically from the allosteric to the active site. 
\item The mouthpiece of the trumpet is surrounded by two more rigid regions, that appear in dark in Fig.~\ref{zp}.A.
\item The coordination number is close to its critical value, called isostatic, in the vicinity of the active site (see below). 
\end{itemize}

The trumpet-like architecture  is robust: it remains qualitatively unchanged as the mean  coordination number is varied, as long as  $\delta z>0$.  For $\delta z<0$ however, a trumpet still exists (see SI), but it is inverted: it is more coordinated than the rest of the system. 

Next we study how such trumpets shape the response to a binding event, by considering the mean-squared magnitude of the normalized response at different nodes $i$, $\langle | \delta{\bf R}_i^r |^2/\sum_i| \delta{\bf R}_i^r |^2\rangle$ as shown in Fig.~\ref{zp}. For random networks, unsurprisingly the response is  large only close to the stimulus site. However, the response of fit networks display a striking feature: it varies {\it non-monotonically } between the allosteric and the active site. It almost vanishes in the bulk of the material, but reappears near the active site where it is the strongest. 


\section*{Physical processes underlying allostery}

The observation that fit networks develop a less-constrained region connecting the stimulus to the active site is not very surprising,
since  the  elastic point response can remain  heterogeneous on longer length scales in that case \cite{Lerner14}. This argument does not explain however the strong asymmetry of the trumpet, more coordinated near the stimulus and nearly marginally connected near the active site. We now argue that this design is selected for because it prevents the decay in the amplitude of the signal one expects in traditional materials.


\begin{figure*}[tbp]
\centering
\includegraphics[height=6.4cm]{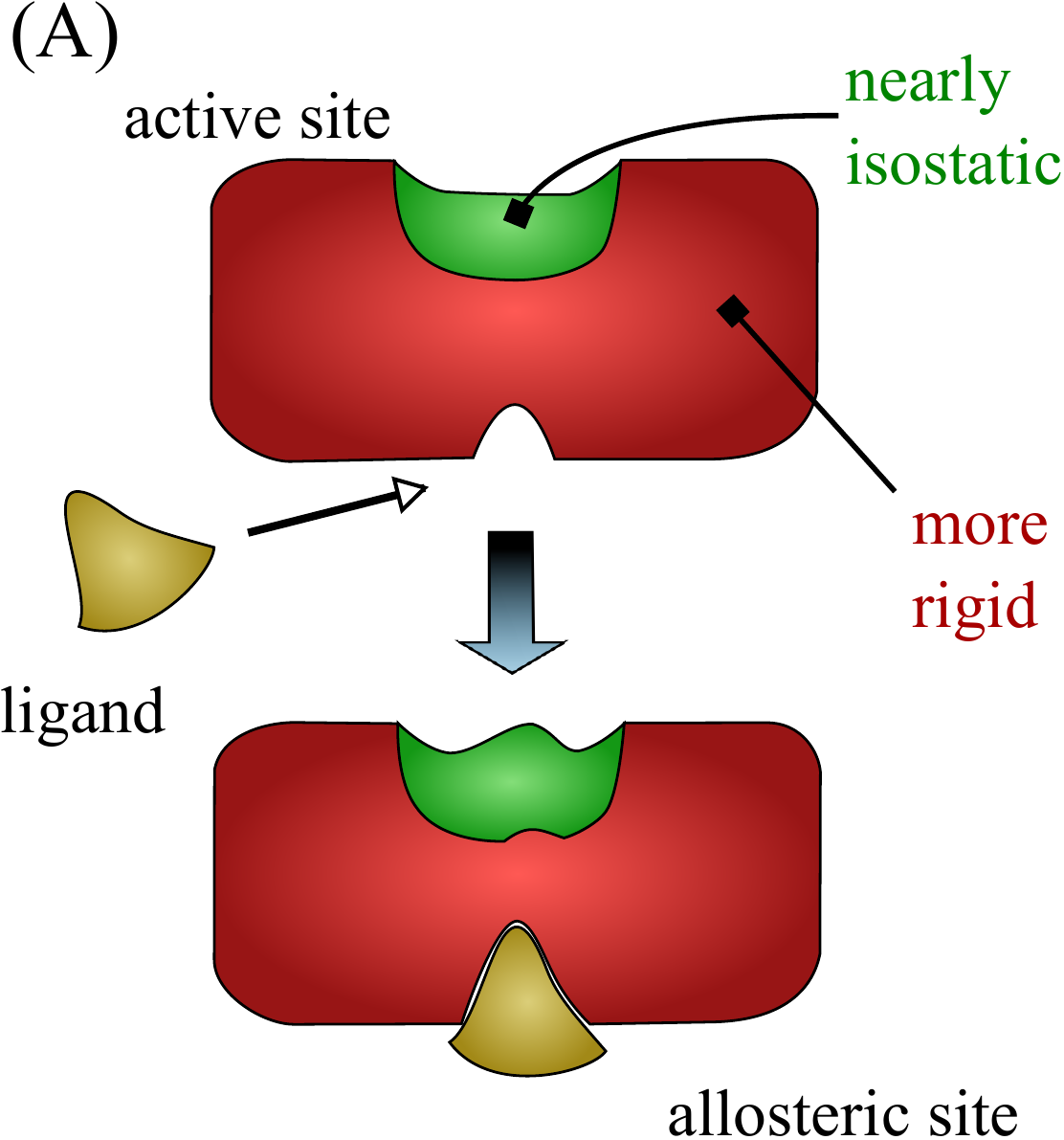}
\includegraphics[height=6.8cm]{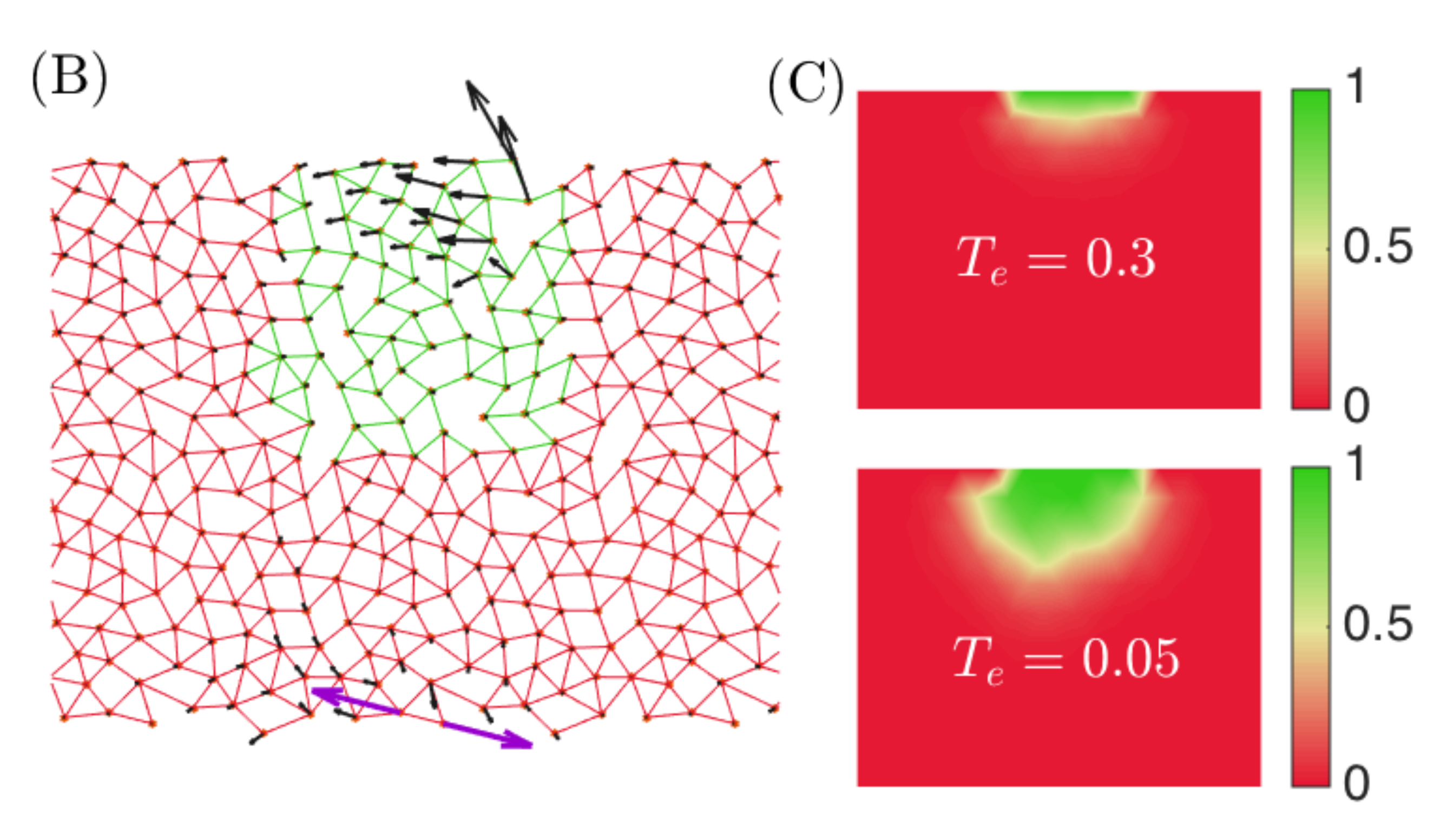}
\caption{
(A) Illustration of the  mechanism responsible for allostery in our artificial networks.They display a nearly isostatic region in the vicinity of the active site, surrounded by a better connected material. When the ligand binds, it induces an effective shape change at the allosteric site. This mechanical signal transmits and decays through the well-connected body of the material. It is then amplified exponentially fast in the isostatic region near the active site, leading to a large strain. (B) System made of two elastic networks with coordination $z=5.0$ (red) and $z=4.0$ (green). Its response (black) to a perturbation (purple) demonstrates that a marginal network embedded in a more connected one can amplify the response near its free boundary. (C) Spatial distribution of the probability that a node is in an isostatic region connected to the active sites at $T_e=0.3$ and $T_e=0.05$ for $z=5.0$.
\label{iso}}
\end{figure*}

Recent works have shown that marginally connected crystals can display edge modes, 
leading to exponentially growing response when displacement are imposed at the boundary of the system 
\cite{Kane14,Chen14,Sussman15}. It was very recently shown that such ``explosive" modes must be present  in disordered marginally connected materials as well \cite{Yan16b}.  Such systems, if sufficiently constrained at some of their boundaries, can act as a lever which can amplify complex motions exponentially toward free boundaries \cite{Yan16b}.  

We argue that our allosteric networks are built along this principle.  
 As sketched in Fig.~\ref{iso}.A., their structure is approximately that of a well-connected elastic material surrounding
 a marginally connected (called ``{\it isostatic}") network near the active site. If a stimulus is imposed on an allosteric site, the response will decay with distance in the well-connected region, leading to imposed displacements of small amplitude on the boundary of the isostatic region. As noted above, ``squeezing"  such systems  leads to an explosive response in the direction of free boundaries, allowing the response to reach the desired amplitude on the target nodes. This explanation captures both the observation that allosteric networks are nearly isostatic near the active site, and that their response varies non-monotonically in space.

To test this proposal we build an artificial network, by embedding a random isostatic network with $z=4$ to a better coordinated random network with $z=5$ (the details of the construction are explained in Methods). As shown in Fig~\ref{iso}.B, the response to an imposed dipole  on the open boundary of the well-connected network decays, but eventually grows rapidly toward the boundary of the isostatic region, as predicted.

As discussed in Methods, we have developed an algorithm to recognize a nearly isostatic region that contains the target nodes in our evolved  networks. In Fig.~\ref{iso}.C we show the probability map that a node belongs to such an isostatic network, obtained by averaging on many realizations.  Fit allosteric networks indeed  show a robust isostatic region attached to the active sites, which is absent for random networks. 

\section*{Conservation and co-evolution}
 In our model, the ``sequence" of a network corresponds to the vector of zeros and ones $|\sigma\rangle$ that defines a structure. This is analogous to the sequence of amino acids that defines a protein. Using our Monte-Carlo at some low $T_e$, we can generate a family of  sequences associated with  networks of high fitness. If only such a  family could be observed (and assuming no knowledge on the task being performed nor on the spatial organization of the networks), would it be possible to infer  which region of the system matters for function?
 There is evidence that such inference is useful for some protein families, if enough sequences are available \cite{Halabi09,mclaughlin12}. We show that it is also the case in our model.

\begin{figure}[h!]
\centering
\includegraphics[height=6.cm]{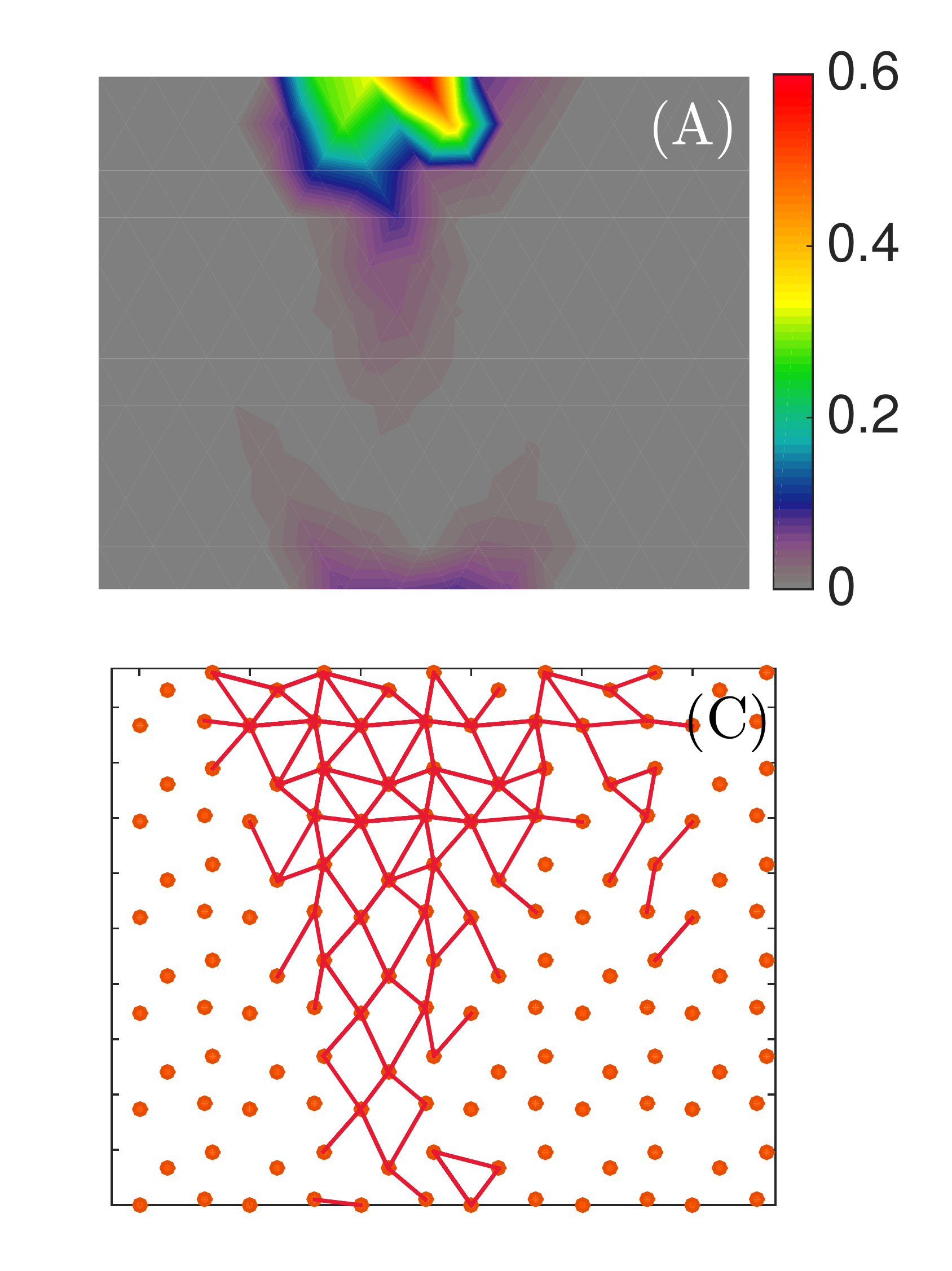}
\includegraphics[height=6.1cm]{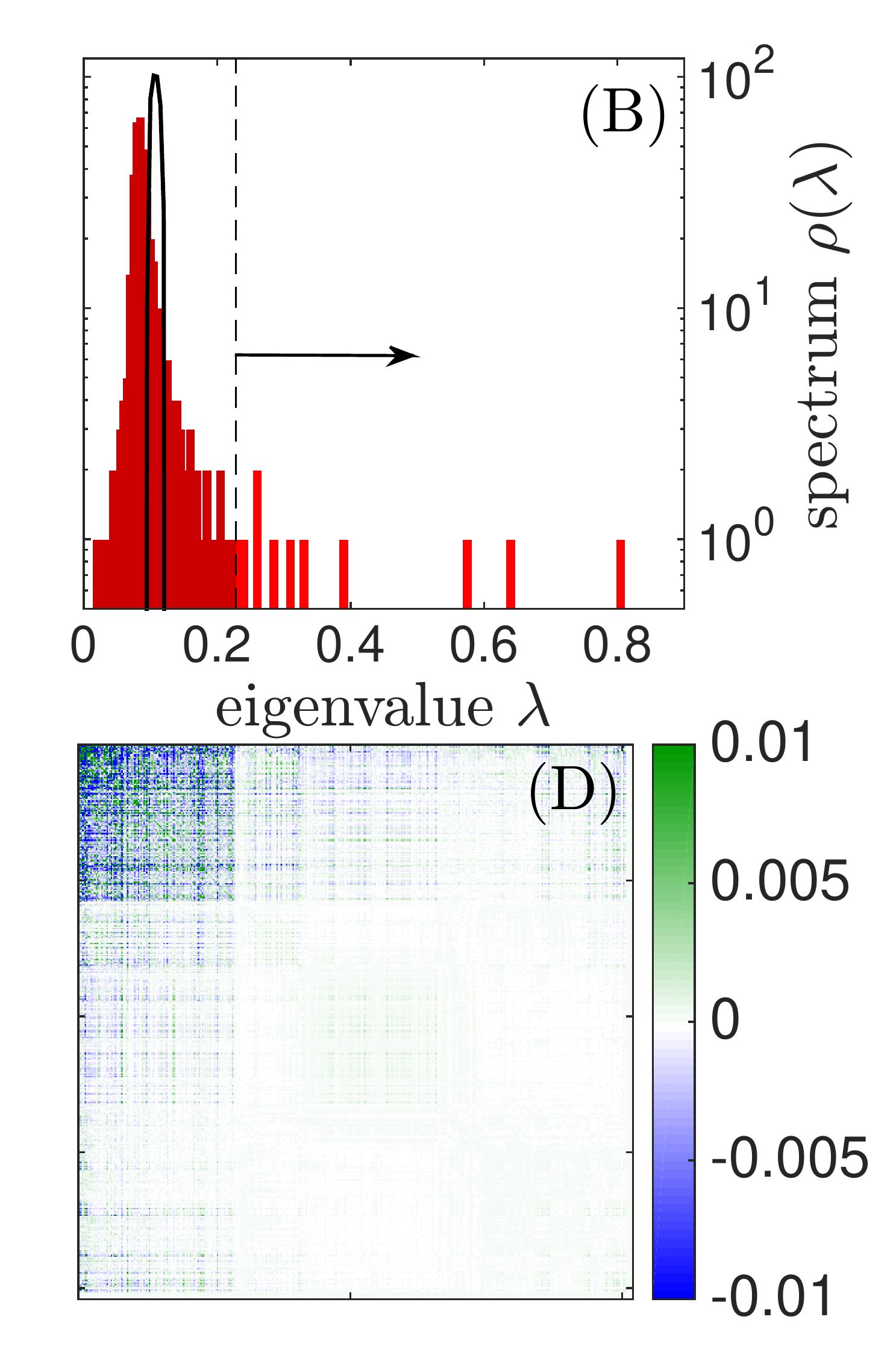}
\caption{(A) Spatial distribution of conservation, as defined in the text, for $T_e=0.05$ and $z=5.0$. { (B)} Spectrum of eigenvalues $\rho(\lambda)$ of $C$ for the high temperature case ($T_e=0.30$) in black and low temperature ($T_e=0.05$) in red. The arrow indicates which eigenvalues are used to identify the springs shown in (C). { (C)} Springs selected using the procedure explained in the text. {(D)} $\tilde{C}$ is built using the same parameters as in (C).  $\tilde{C}$ presents  a clear separation in a region where the correlations are stronger, which corresponds to the trumpet shown in (C). All these figures are made using $L=12$ and $z=5$.}
\label{sector}
\end{figure}

Key aspects of the design are more likely to stay conserved in evolution. Here we define conservation  in  each link $\alpha$ as \cite{Cover06}:
\be
\Sigma_\alpha\equiv \langle\sigma_{\alpha}\rangle\ln\frac{\langle\sigma_\alpha\rangle}{\bar\sigma}+(1-\langle\sigma_\alpha\rangle)\ln\frac{1-\langle\sigma_\alpha\rangle}{1-\bar\sigma},
\label{consv}
\ee
where $\langle\sigma_\alpha\rangle$ is the ensemble average of the occupancy of link $\alpha$ and $\bar\sigma=N_s/(3N-2L)$ is the mean occupancy of the links. $\Sigma_\alpha$ is a measure of the predicability of the occupancy of the link $\alpha$: it  is zero when the link occupancy is random, and maximum when the link is always  empty or always occupied. 
The conservation map of allosteric networks is shown in Fig.~\ref{sector}.A. We can distinguish the trumpet pattern, but  most strikingly, the neighborhood of the active site is very conserved.  This observation supports that  specificity of the response is essentially controlled by the geometry of the network near the active site. 
  
Next, we test if an analysis of co-evolution alone reveals important features of function and structure. 
A more quantitative discussion will be presented elsewhere, here we highlight the main results.
We define the correlation matrix $C$ between the links $\alpha,\beta$ as:
\begin{eqnarray}
  C_{\alpha\beta} =  \langle \sigma_\alpha \sigma_\beta \rangle - \langle \sigma_\alpha \rangle \langle \sigma_\beta \rangle,
  \label{Cij}
\end{eqnarray}
where the $\langle\bullet \rangle$ is again the ensemble average over the solutions found by the Monte Carlo algorithm. We then compute the  eigenvalues $\lambda_1>\lambda_2 > ... > \lambda_{N_{\rm s}}$ of the matrix $C$.  Fig.~\ref{sector}.B compares the  spectrum of eigenvalues of a high temperature (essentially random) network with that of allosteric networks obtained at small $T_e$. In the latter case, some eigenvalues are much larger than the continuum spectrum, itself much more spread than in the random case. 

To select out a ``sector" \cite{Halabi09} of links that co-evolve, we: (i) pick up the $N_{\Gamma}=10$ eigenvectors $|\psi^{\gamma} \rangle$ with highest eigenvalues; (ii)  we include a given link $\alpha$ in the sector if for at least one of this ten modes, $|\psi^{\gamma}_\alpha|>\epsilon=0.05$.  Links selected in this procedure are shown in Fig.~\ref{sector}.C. This procedure selects precisely the links that belong to the trumpet, supporting the idea that co-evolution alone can uncover key functional aspects \cite{Halabi09,mclaughlin12}.


For completeness,  we define a correlation matrix reconstructed from its 10 top eigenvectors: 
\begin{eqnarray}
  {\tilde{C}_{\alpha\beta}}= \sum_{\gamma=1}^{N_{\Gamma}} \lambda_{\gamma} |\psi^{\gamma} \rangle \langle \psi^{\gamma}|.
\label{Cij2}
\end{eqnarray}
 ${\tilde{C}}$ is shown in Fig.~\ref{sector}.D  after re-ordering links in terms of the strength of their components in the top ten modes, clearly showing a sector of links where correlations are strong in amplitude (but vary in sign).



\section*{Conclusion}

We have introduced a scheme to discover materials that  accomplish a specified task. It allows us to characterize the architecture of the solutions, their entropy and how correlated they are. We illustrated this approach using a specific allosteric task, where a strain imposed on an allosteric site must lead to a given strain on a distant active site. The architectures we obtain are highly anisotropic. Our analysis revealed that the physical mechanisms that enable allostery include the recently discovered presence of soft edge modes in weakly-connected elastic materials \cite{Yan16b}. It would be very interesting to test if some  proteins have evolved to exploit such surface effects to operate. In addition, our analysis supports the notion \cite{Halabi09,mclaughlin12} that analyzing the correlation of the "sequences" defining structure can indeed reveal compact regions  central for function.  

The detailed study of the architectures we found also reveals some of their limitations. 
Real proteins have additional constraints than those we have considered: among others, they are made of a chain that folds and remains relatively stable despite thermal noise. Our asymmetric structures are quite soft near the active site: as documented in S.I, the thermally-induced motion  would be about four times larger there than in the other nodes also located at the system surface, which may not be desirable. Note however that such features  will   improve if alternative fitness functions are considered, which our approach allows for. This could be implemented by  explicitly penalizing thermal motion at the active site. An intriguing  
extension of our work is to reason in terms of energy, instead  of displacement,  and  maximize the {\it cooperativity} between two distant sites. Denoting by $E_1$ and $E_2$  the mechanical energies associated with a binding event in some site 1 and site 2 respectively, and $E_{12}$ the energy of binding both, we can consider $F=E_1+E_2-E_{12}$. Fitness can be large only if the two sites are strongly coupled together elastically, which from the symmetry of the fitness presumably corresponds to more symmetric architectures than those discovered here. 


\section*{Methods}
\subsection*{Computing the linear response to an imposed displacement.}
The linear response to an external force field  $|{\bf F}\rangle$ reads:  
\be
|{\bf F}\rangle=\mm|\delta{\bf R}\rangle
\label{fmr}
\ee
where the stiffness matrix $\mm$ depends only on connection $|\sigma\rangle$ and the link directions. 

To impose the stimulus at the allosteric nodes $|\delta{\bf R}^{\cal E}\rangle$, forces must be applied on these nodes. All other nodes adapt to a new mechanical equilibrium with no net forces on them, and follow a displacement $|\delta{\bf R(\sigma)}^r\rangle$. Thus Eq.[\ref{fmr}] becomes:
\be
\lp\begin{array}{c}|{\bf F}^{\cal E}\rangle\\|{\bf 0}\rangle
\end{array}\rp
=\mm
\lp\begin{array}{c}
|\delta{\bf R}^{\cal E}\rangle\\
|\delta{\bf R(\sigma)}^r\rangle
\end{array}\rp.
\ee 
which leads to:
\be
\lp\begin{array}{c}
|{\bf F}^{\cal E}\rangle\\
|\delta{\bf R(\sigma)}^r\rangle
\end{array}\rp
={\cal Q}^{-1}\mm
\lp\begin{array}{c}
|\delta{\bf R}^{\cal E}\rangle\\
|{\bf 0}\rangle
\end{array}\rp
\ee
with 
\be
{\cal Q}_{ij} = \left\{
\begin{array}{cc}
\delta_{ij}& {\rm if} j\in{\cal E}\\
\mm_{ij}& {\rm if} j\not\in{\cal{E}}
\end{array}\right.
\ee

\subsection*{Computing the fitness.}
Minimizing  Eq.(\ref{cost}) with respect to the global translation and rotation leads to:
\be
\sqrt{\sum_{i\in{\cal T}}(\delta{\bf R}^r_i-\delta{\bf R}^{\cal T}_i)^2-\sum_{i\in{\cal T}}(\delta{\bf R}^r_i-\delta{\bf R}^{\cal T}_i)(\delta{\bf R}^r_{i+1}-\delta{\bf R}^{\cal T}_{i+1})}
\ee
where $i+1=\min(\cal T)$ if $i=\max(\cal T)$.

\subsection*{Metropolis algorithm.}
Starting from a configuration $|\sigma\rangle$, we consider the move toward a new configuration $|\sigma'\rangle$ that differs only by the motion of a strong spring. The move is accepted with the probability: 
\be
P(|\sigma\rangle\to|\sigma'\rangle)=\min[1,\exp\lp\frac{F(|\sigma'\rangle)-F(|\sigma\rangle)}{T_e}\rp],
\label{detailed}
\ee
which satisfies detailed balance. This finishes one Monte Carlo step.  The next step starts from the configuration just sampled.

For each coordination number $z$ and each evolution temperature $T_e$, we sample $20$ Monte Carlo sampling series with $10^5$ Monte Carlo steps in each, and do not consider the first half of these time series (which is sufficient to eliminate transient effects). Our results are thus averaged over $10^6$ configurations. 

\subsection*{Mosaic network.}
To construct the mosaic network of Fig.~\ref{iso}B, we first generate a highly coordinated random network by annealing $N$ bidisperse soft discs with radius ratio $1.4:1$ to mechanical equilibrium. We  then produce networks with different coordination numbers by removing springs between the most connected neighbors one by one~\cite{During13,Yan16}. This procedure ensures that the spatial fluctuations of coordination are low. 

We then cut the periodic boundaries in  two networks, with $z=4.0$ with $z=5.0$ respectively, obtained from the same packing (and thus presenting nodes at the same locations). The network with $z=5.0$ is duplicated three times. We complete a four square made of these patches, and connect nodes through the patch boundaries. We apply a periodic boundary condition in the horizontal direction. 

\subsection*{Algorithm to detect the isostatic region around the active site.}
The pebble game algorithm~\cite{Jacobs95} used to identify  independent rigid clusters cannot reveal a nearly isostatic network embedded in an over-constrained region. Here we introduce an algorithm to do that. The basic idea is to search for a connected set of nodes including the active site where Maxwell's rigidity condition~\cite{Maxwell64} is  satisfied to a desired level of accuracy. Specifically, we: \\ 
({\it i}) Initialize the searching subnetwork with the target nodes. The subnetwork includes the nodes inside, the internal connections between these nodes and the external connections to the nodes outside of the subnetwork. \\
({\it ii}) Count the degrees of freedom $N_f$ and the number of constraints $N_c$ in the subnetwork. $N_f=d n_0$, where $n_0$ is the number of nodes inside the subnetwork.  $N_c=n_I+q n_E$, where $n_I$ and $n_E$ are the numbers of internal and external connections. The parameter $q\in[0,1]$.\\ 
({\it iii}) If $N_f>N_c$, compute $N_f'$ and $N_c'$ for each external node linked by an external connection when including it into the subnetwork.\\
({\it iii'}) Otherwise, stop.\\
({\it iv}) Update the subnetwork by including the nearest node with minimal $N_c'-N_f'$ and set $N_f=N_f'$ $N_c=N_c'$. \\
({\it v}) Check and include the nodes all of whose neighbors are in the subnetwork. Go to step ({\it iii}). 

The algorithm identifies a compact region depending on the parameter $q$ that counts the external connections. In the results presented in Fig.~\ref{iso}.C, we have chosen $q=0.5$ ($q=1$ corresponds to the pebble game). We discuss the physical meaning of different $q$ and show that $q=0.5$ is an appropriate choice in SI.


\begin{acknowledgments}
We thank B. Bialek, J-P Bouchaud, P. De Los Rios, E. DeGiuli, D. Malinverni, R. Monasson, O. Rivoire and  S. Zamuner for discussions. L.Y. was supported in part by the National Science Foundation under Grant No. NSF PHY11-25915.  M.W. thanks the Swiss National Science Foundation for support under Grant No. 200021-165509 and the Simons Foundation Grant ($\#$454953 Matthieu Wyart). {This material is based upon work performed using computational resources supported by the ``Center for Scientific Computing at UCSB'' and NSF Grant CNS-0960316. }
\end{acknowledgments}


\bibliographystyle{unsrt}
\bibliography{Wyartbibnew}
\newpage

\renewcommand{\theequation}{S\arabic{equation}}
\setcounter{equation}{0}
\setcounter{figure}{0}
\renewcommand{\thefigure}{S\arabic{figure}}

\subsection*{Supplementary Information (SI)}

\subsubsection*{A. Distortion of triangular lattice}
\begin{figure}[htbp]
\centering
\includegraphics[width=6.5cm]{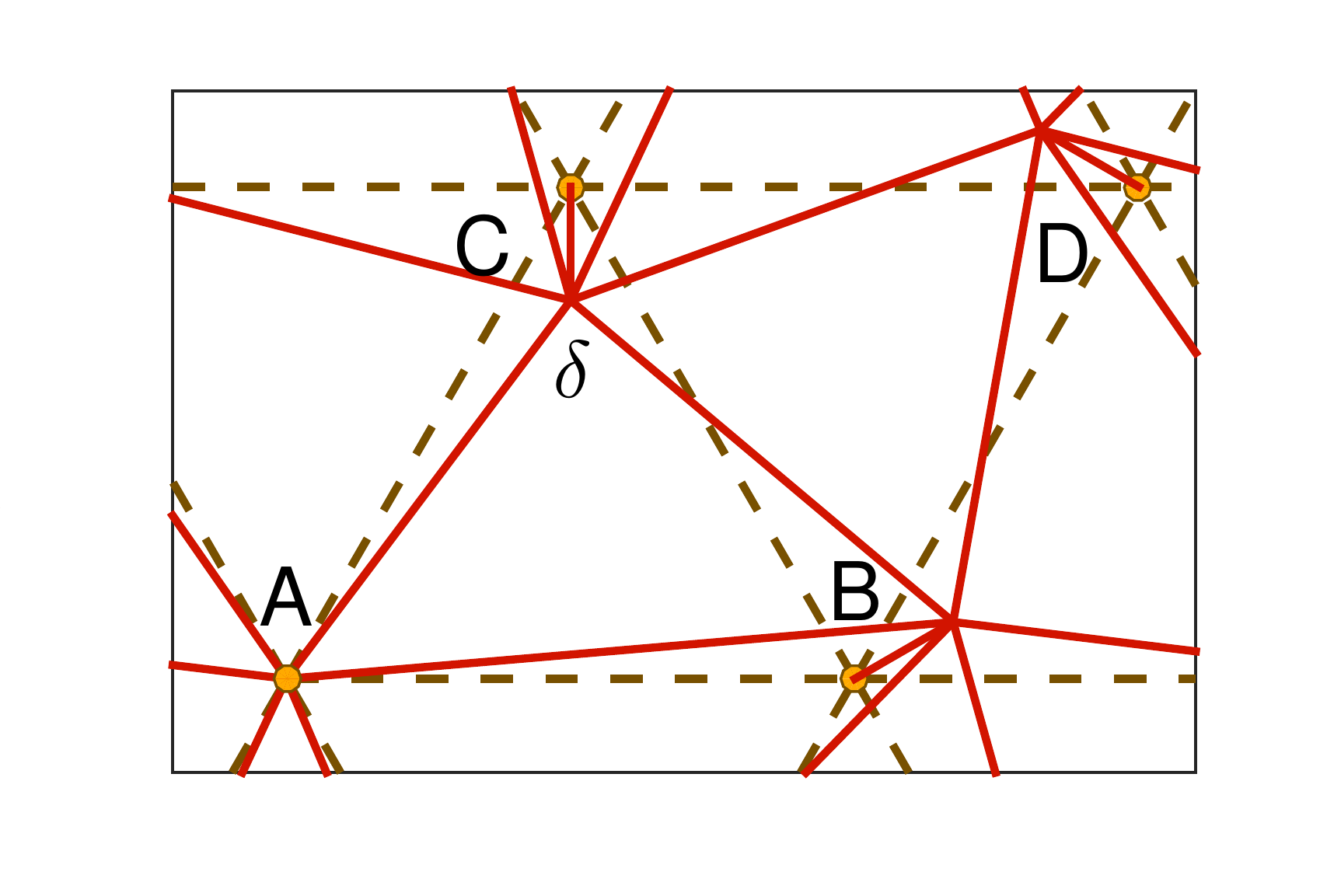}
\caption{Illustration of the distorted triangular lattice. 
\label{distortion}}
\end{figure}

In our model, we introduce a slight distortion of the lattice to remove long straight lines that occur in a triangular lattice. Such straight lines are singular and lead to unphysical localized floppy modes orthogonal to them.  
One can remove them by imposing a random displacement on the nodes. Instead, we distort the lines without introducing frozen disorder. We group nodes in lattice by four, labeled as A B C D in Fig.~\ref{distortion}. 
One group forms a cell of our distorted lattice. In each cell, node A stays in place,  while nodes B, C, and D move by some distance $\delta$: B along the direction perpendicular to BC, C along the direction perpendicular to CD, and D along the direction perpendicular to DB, as illustrated.  We set $\delta$ to  $0.2$, where the straight lines are maximally reduced with this distortion. 

\begin{figure}[htbp]
\centering
\includegraphics[width=1.\columnwidth]{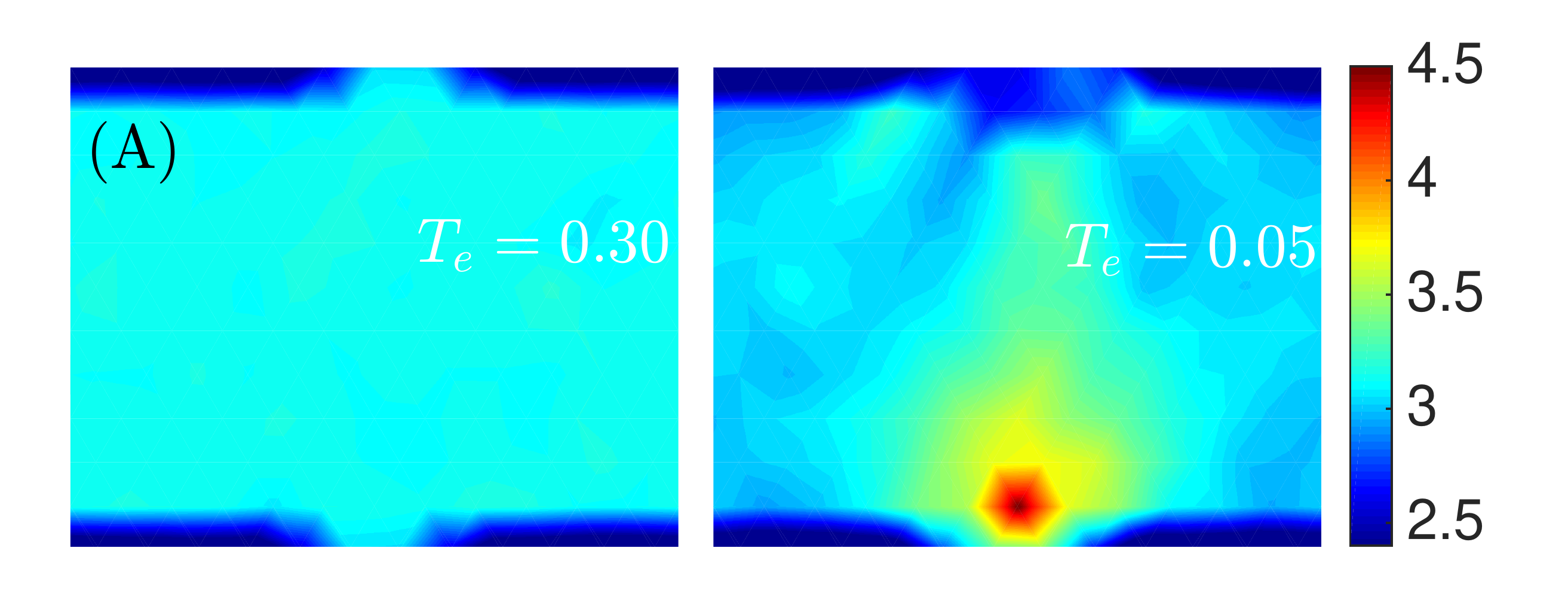}\\
\includegraphics[width=1.\columnwidth]{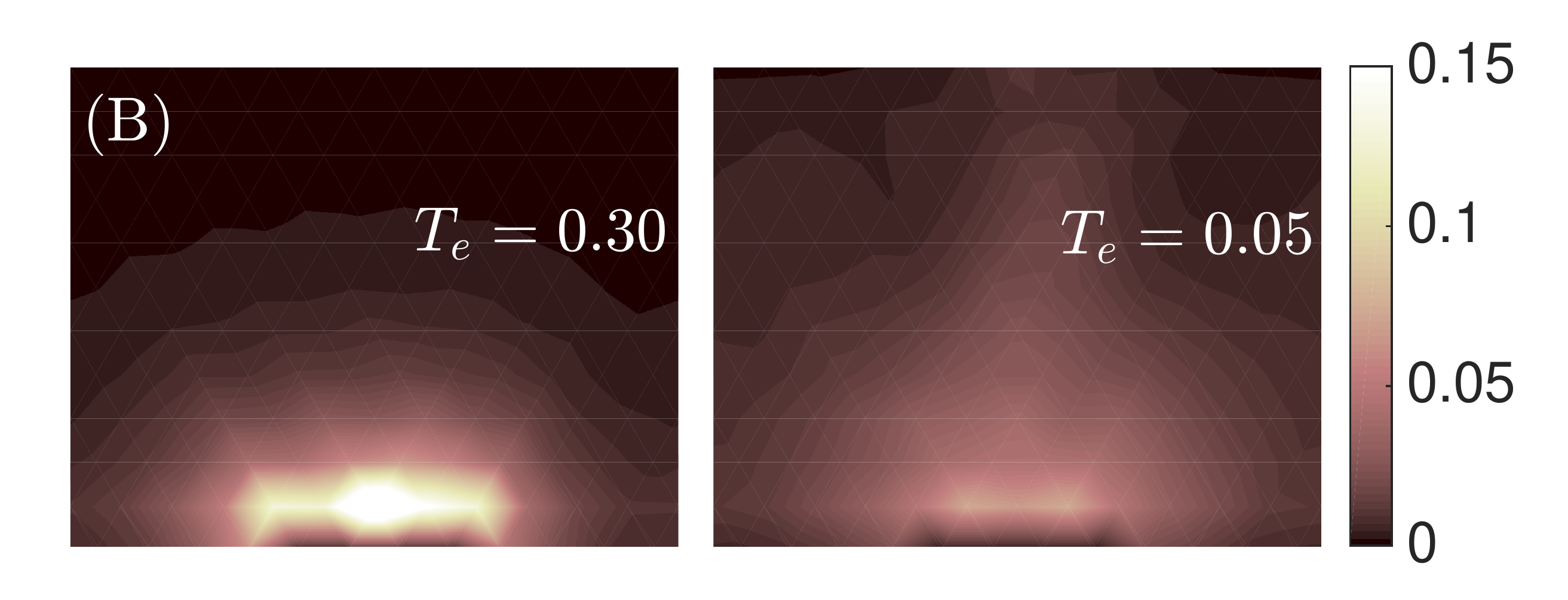}
\caption{A. Spatial distribution of coordination number. B. Spatial distribution of the response magnitude to the excitation at allosteric sites. Left: $T_e=0.3$. Right: $T_e=0.10$. $z=3.0$. 
\label{floppy}}
\end{figure}

\subsubsection*{B. ``Trumpet'' structure in floppy networks}

In general, the displacement signal can only propagate a finite distance in rather homogeneous floppy networks \cite{During13}. 
This can be seen in the left panel of Fig.~\ref{floppy}.B. 

Below the transition temperature ($T_c=0.12$ for $z=3.0$), we find that the network resolves this issue by generating a ``trumpet'' shape structure similar to the rigid case, connecting the allosteric site and the active site, as appears in  the average coordination map shown in Fig.~\ref{floppy}.A. The coordination number in this trumpet structure is  larger than in the rest of the material, allowing the signal to propagate. Similarly to the $\delta z>0$  case, the mean coordination number decreases monotonically from the allosteric to the active site. However,  the trumpet has the wider nearly-isostatic patch near the allosteric side. We also find that the amplitude of the displacement decreases monotonically from the allosteric to the active side in the bottom panel of Fig.~\ref{floppy}.B. 


\subsubsection*{C. Thermal noise}
We quantify thermally-induced motion in a structure by the B-factor~\cite{Trueblood96} (related to the Debye-Waller factor), 
\be
B=8\pi^2\langle u^2\rangle
\ee
where $u$ is the magnitude of particle displacement around its mean position, and $\langle\bullet\rangle$ denotes thermal averaging. 

\begin{figure}[htbp]
\centering
\includegraphics[width=1.0\columnwidth]{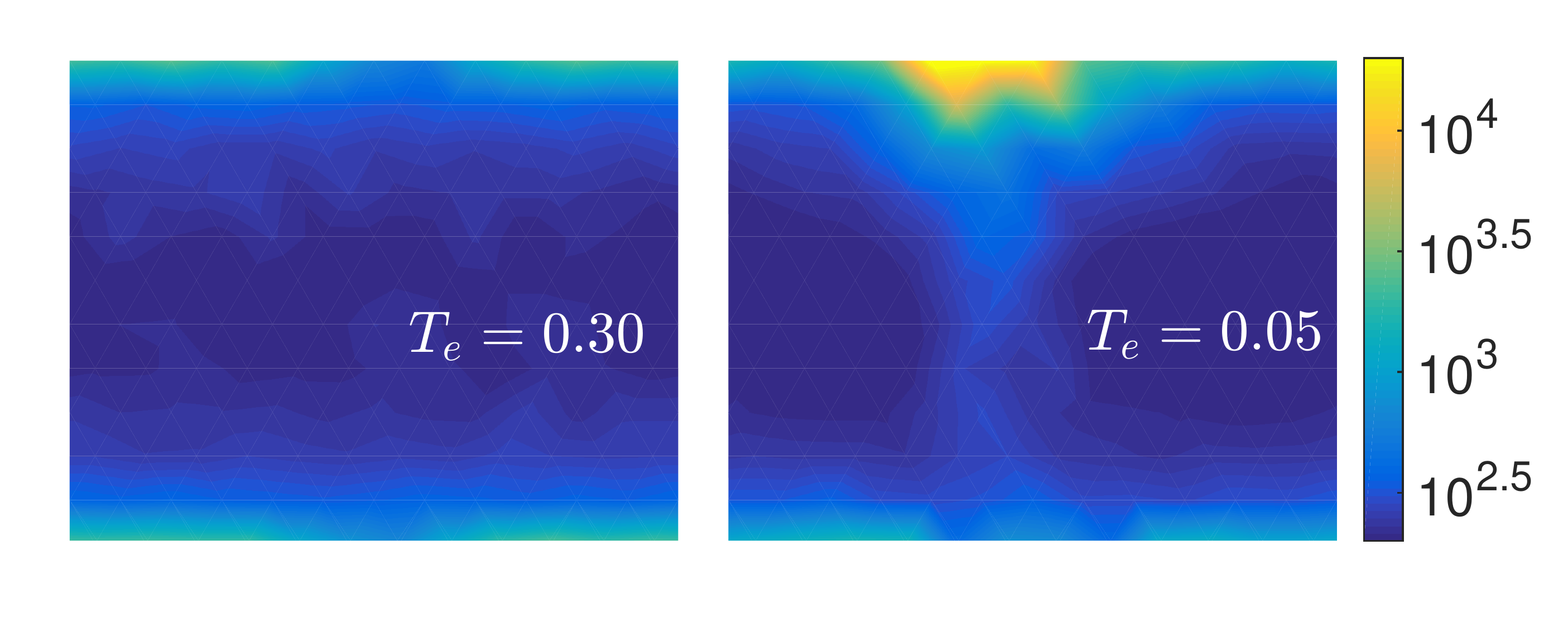}
\caption{Spatial distribution of B-factor for $T_e=0.30$ (left) and $T_e=0.05$ (right). Colorbar is in log scale, $z=5.0$.
\label{bfac}}
\end{figure}

In an elastic network, the energy due to displacement field $\vec{u}_i$ is,
\be
\mh=\frac{1}{2}\sum_{i,j}\vec{u}_i\cdot\mm_{ij}\cdot\vec{u}_j
\ee
where $\mm$ is the stiffness matrix defined in Eq.(\ref{fmr}). The B-factor is directly related to the stiffness matrix $\mm$, as we now recall. 
By definition, the thermal average of the correlation of the displacement at temperature $T$ is
\be
\langle \vec{u}_k\cdot\vec{u}_l\rangle(T)=\frac{1}{Z(T)}\int\prod_i\rd\vec{u}_i\vec{u}_k\cdot\vec{u}_le^{-\frac{1}{2T}\sum_{i,j}\vec{u}_i\cdot\mm_{ij}\cdot\vec{u}_j},
\ee
a Gaussian integral. It can thus be carried out, 
\be
\langle \vec{u}_i\cdot\vec{u}_j\rangle=T\lp\mm^{-1}\rp_{ij},
\ee
and
\be
B_i=8\pi^2\langle{u}_i^2\rangle=8\pi^2T(\mm^{-1})_{ii}.
\ee
We compute B-factor with temperature $T=1$. 

In Fig.~\ref{bfac}, we show the B-factor map for the configurations at high and low evolution temperature. Thermal motions are stronger in the``trumpet'' shape discussed in the main text, especially at the vicinity of the active site. There, the B-factors are about $20$  times larger than in the other nodes at the boundary of the system, corresponding to an amplitude of motion about four times larger. 

\begin{figure*}[htbp]
\centering
\includegraphics[width=15cm]{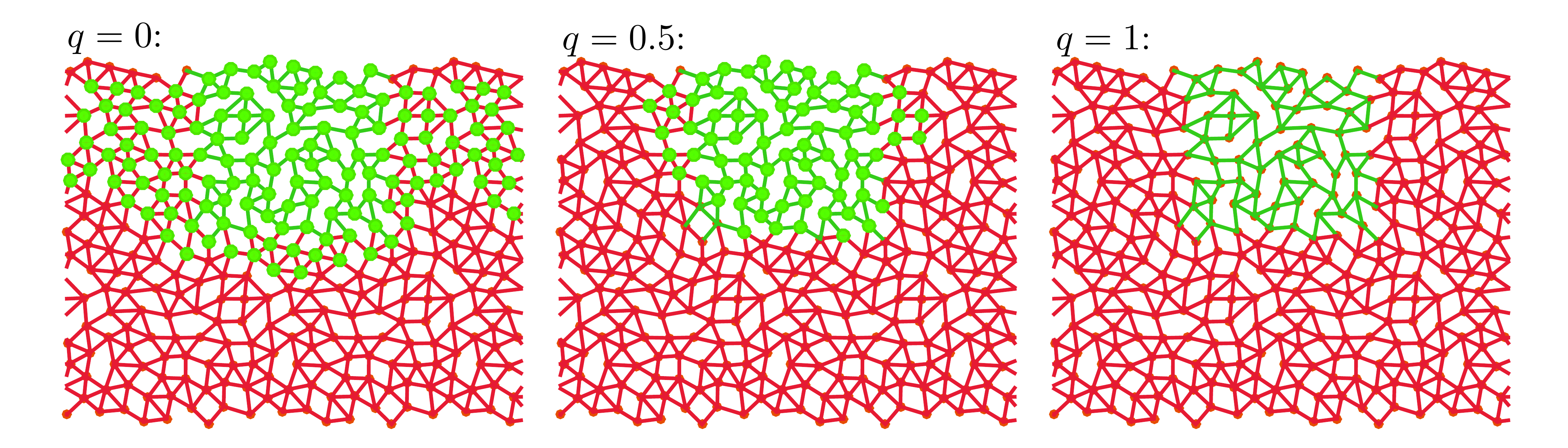}
\caption{Isostatic region identified by the constraint-counting algorithm for the mosaic network. Identified region is shown by green nodes. Green and red lines are links belonging to $z=4.0$ and $z=5.0$ patches in the mosaic network. From left to right, external constraints are counted as $q=0$, $q=0.5$, $q=1$.
\label{isoiden}}
\end{figure*}

\subsubsection*{D. Counting of the external constraints}

We have described an algorithm to detect the nearly isostatic region near the target nodes, in which there is an undetermined parameter $q$, the fraction of counting an external constraint. 
$q$ describes the boundary condition of the subnetwork: $q=0$ for open boundary condition; $q=1$ for fixed boundary condition; $q=0.5$ for a boundary equivalent to the bulk. We find that $q=0.5$ works better than $q=0$ or $q=1$ numerically in identifying the nearly-isostatic region for the mosaic network, shown in Fig.~\ref{isoiden}.
\end{article}

\end{document}